\documentclass[fleqn,twoside,twocolumn,nofootinbib,showkeys]{revtex4} 
\usepackage[sec,nocpr]{ujp} 

\begin{document}
	\title[]
	{ PION PRODUCTION IN $\nu_\mu$ CHARGE CURRENT INTERACTIONS ON $^{40}Ar$ IN DEEP UNDERGROUND NEUTRINO EXPERIMENT}%
	\author{Ritu Devi}
	\affiliation{Department of Physics, University of Jammu, Jammu, India}
	\address{Jammu Tawi-180006}
	\email{rituhans4028@gmail.com}
	\author{Baba Potukuchi}
	\affiliation{Department of Physics, University of Jammu, Jammu, India}
	
	\udk{¹ ÓÄÊ/UDC} \razd{\secix}
	
	\autorcol{R.\hspace*{0.7mm} Devi, B.\hspace*{0.7mm} Potukuchi.}%
	
	\setcounter{page}{1}%

\begin{abstract}
Understanding pion generation and the consequences of final state interactions (FSI) is critical for data processing in all neutrino experiments. The energy utilised in modern neutrino research, the resonance (RES) generation process contributes significantly to pion production. If a pion is absorbed in the nuclear matter after it is produced, the event may become unrecognizable from a quasi-elastic (QE) scattering process and act as a background. For oscillation experiments, estimating this background is critical, and it necessitates solid theoretical models for both pion generation at the primary vertex and after FSI. The number of final state pions created differs greatly from the number produced at the primary vertex due to FSI. Because neutrino detectors can only detect final state particles, FSI obscures the proper information about particles created at the primary vertex. A detailed study of FSI is required to overcome this problem, which theoretical models incorporated in Monte Carlo (MC) neutrino event generators can provide. They provide theoretical results of neutrino interactions for diverse research, acting as a connection among both theoretical models and experimental data.  In this paper, we provide simulated events for pion creation in $\nu_\mu$ charge current (CC) interactions on a $^{40}$Ar target in the Deep Underground Neutrino Experiment (DUNE) experimental setup for two distinct MC generators-GENIE and NuWro. In comparison to GENIE (v-3.00.06), NuWro (v-19.02.2) is more opaque (less responsive) to charge exchange and absorption processes; pions are more likely to be absorbed than produced during intranuclear transport.

\end{abstract}


\maketitle

\section{Introduction}
Neutrino physics is reaching a precise era, powered by new experiments and contemporary detector technology, and this necessitates a better theoretical and phenomenological explanation of neutrino interactions. Neutrinos rarely interact with matter and can move rapidly without interacting with it. The properties of neutrinos are yet unknown, making their study difficult from both a theoretical and an experimental standpoint. The neutrino interactions are described by the Standard Model (SM)'s electroweak theory. SM assumed the neutrino was a massless particle in its previous formulation, hence mass mixing was not expected, unlike quarks.  If neutrinos had mass, mass mixing would be possible in the lepton sector, and a neutrino generated in one flavour could later be seen as a neutrino of some other flavour, a phenomenon known as neutrino oscillations. The masslessness of neutrinos is not required by any fundamental physics principle. The Pontecorvo-Maki-Nakagawa-Sakata (PMNS) matrix \cite{pmns} defines the strength of mass mixing in the leptonic sector. Three mixing angles ($\theta_{12}, \theta_{23}, \theta_{13}$), and the CP phase factor ($\delta_{CP}$) can be used to represent PMNS. Apart from these mass mixing factors, the probability of neutrino oscillation is determined by the mass of neutrinos (or the difference of their squares), i.e. $\Delta m^2_{21} = m^2_2 - m^2_1 ; \Delta m^2_{32} = m^2_3-m^2_2 ; \Delta m^2_{31} = m^2_{32} + m^2_{21} $.

Understanding CC neutrino-nucleus interactions in the few GeV energy region is critical for current and future neutrino experiments. However, understanding neutrino-nucleus cross-sections, modelling of hadronization and intranuclear hadron transport,  description of nuclear models, and nuclear effects in this energy region is difficult and requires many intermediate steps. For this, we will need a conventional MC generator that can account for all of these phases. GENIE \cite{genie}, ANIS \cite{anis},  NuWro \cite{nuwro}, GiBUU \cite{gibuu}, MARLEY \cite{marley}, NEUT \cite{neut}, and Nuance \cite{nuance} are some of the MC generators are devoted to the modeling of neutrino interactions. All of these generators have the same assumptions. Each generator considers primary  and final state neutrino interactions  individually.

GENIE and NuWro generators are used in this study to simulate pion generation in neutrino-nucleus interactions. Pions are a common backdrop \cite{abratenko} in so many oscillation studies and their processes in FSI \cite{naaz} make them theoretically problematic. DUNE flux has been used for interactions on the $^{40}Ar$ target in both GENIE and NuWro. Only the CC interactions were analysed for each generator. Quasi-elastic (QE) scattering, deep-inelastic scattering (DIS), resonance (RES) production,  and coherent (COH) pion generation were enabled processes. After that, the data was examined for different pion topologies before (primary interactions) and after final state interactions (secondary interactions). Understanding neutrino-nucleus interactions are critical for current experiments like T2K \cite{t2k} and NOvA \cite{nova} as well as upcoming long-baseline neutrino-oscillation experiments like DUNE \cite{DUNE1,DUNE2,DUNE3} and Hyper-KamioKande \cite{Kamiokande}.

The Deep Underground Neutrino Experiment (DUNE) is a global initiative to build a long-baseline neutrino oscillation experiment at Fermi National Accelerator Laboratory (FNAL) in the United States. It has a Near Detector (ND) located 575 metres from the neutrino source and 60 metres underground \cite{DUNE4} at Fermilab in Illinois and a Far Detector (FD) located approximately 1.5 kilometres underground at Sanford Underground Research Facility (SURF) in South Dakota, is nearly 1300 kilometres away from Fermilab. The primary scientific goals of this cutting-edge detector are to conduct a thorough programme of neutrino oscillation observations using Fermilab's $\nu_\mu$ and $\bar{\nu}_\mu$ beams, as well as to restrict the CP violation phase in the leptonic sector. Because the distance between FD and ND are approximately 1300 kilometres, it will give a 1300-kilometer baseline facility for studying matter effect. Both ND and FD will observe the neutrino spectrum with Ar target material, which will enable to overcome numerous systematic uncertainties. The unoscillated neutrino spectrum will be observed by ND, whereas the oscillated neutrino spectrum will be observed by FD. The proportions and technology of the ND and FD at DUNE will be different. The DUNE Near Detector is made up of three basic detector components, two of which can travel off the beam axis: 1) A 50-ton LArTPC (ND-LAr) with pixellated readout built with ArgonCube. 2) The ND-GAr detector, which consists of a high-pressure gaseous argon TPC enclosed by an electromagnetic calorimeter (ECAL) in a 0.5 T magnetic field. 3) The System for on-Axis Neutrino Detection (SAND), an on-axis beam monitor that tracks neutrino flux. It comprises of a huge solenoidal magnet with an inner tracker surrounding by an ECAL. In the current scenario, two inner tracker methods are being considered: one using a mix of plastic scintillator cubes and TPCs, and the other using straw tubes.

\begin{figure}[hbt!]
	\vskip1mm
	\includegraphics[width=\column]{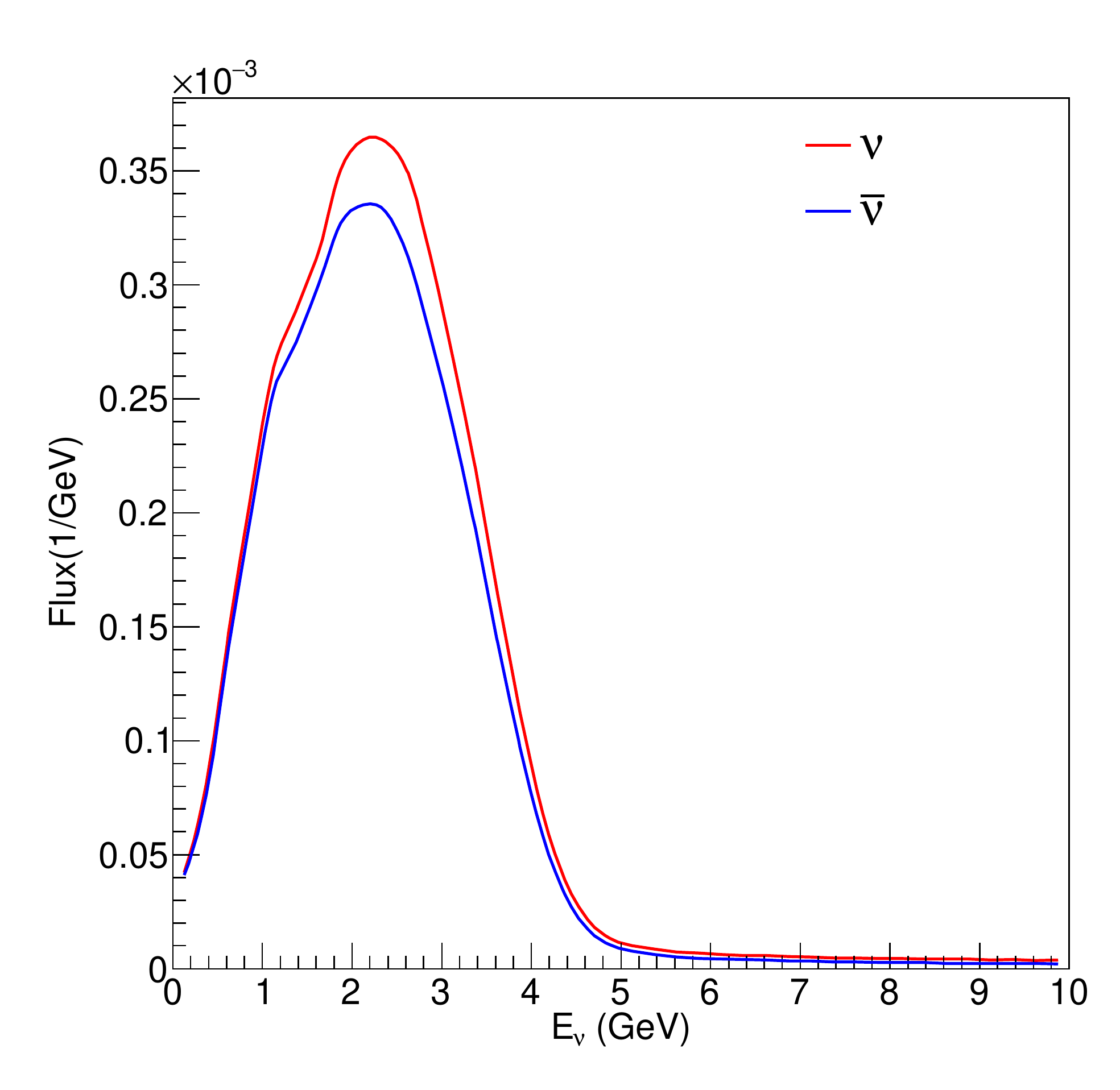}
	\vskip-3mm\caption{The DUNE flux as a function of neutrino energy used in our work.  }
	\label{fig1}
\end{figure}

The DUNE neutrino flux \cite{flux} in the range of energies 0.125-10 GeV was employed in our simulation.
Figure 1 depicts the $\nu_\mu$ and $\bar{\nu}_\mu$  flux  employed in our simulations. It  peaks at about 2.5 GeV and covers the energy range from a few hundred MeV to tens of GeV. The Neutrino at Main Injector (NuMI) beamline facility at Fermilab produces  high purity, an intense, wide-band neutrino beam with an initial output of 1.2 MW (which will be expanded to 2.4 MW) that is estimated to produce $1.1\times10^{21}$ protons each year. The primary beam of protons from the main injector accelerator, with energies ranging from 60 to 120 GeV, collides with the graphite target, producing pions and kaons. With the help of magnetic horns, these mesons will be focused even further toward a 200-meter-long decay pipe, in which they will break into neutrinos and leptons. By reversing the polarity of focussing magnets, the neutrino, and anti-neutrino beams can be expelled individually.

The portions of this paper are as follows: The pion generation in neutrino-nucleus interactions is described in Section II. The processes of QE, RES, DIS, and COH are explained in depth in this section. Section III provides an overview of the GENIE and NuWro MC generators, as well as the numerous models that they employ. The simulation results are presented in Section IV, followed by a summary and conclusions in Section V.

\section{Pion production in neutrino-nucleus interactions}

The most basic account of neutrino-nucleus interactions consist of a description of neutrino-nucleus scattering and a framework for nucleons in the nucleus. Modeling neutrino-nucleus interactions are complicated due to the need to combine many different ideas. We will focus on neutrino-nucleus interactions in the nuclear context. Generator-specific hadronization process explanations and final state interaction models are common. 

 In its simplest and most popular form (known as the impulse approximation technique), neutrino-nucleus scattering is defined as the incoherent sum of scattering from unbound nucleons in the nucleus. However, because nucleons in the nucleus are bound states rather than independent particles, finding the cross-section necessitates a more complex understanding of nuclear dynamics. Charged-current neutrino-nucleus scattering has a total cross-section of \cite{kuzmin}: 
 
 \begin{equation}
 \sigma^{tot}_{\nu N} = \sigma^{QE}_{\nu N}+\sigma^{1\pi}_{\nu N}+\sigma^{2\pi}_{\nu N}+........ \sigma^{1K}_{\nu N}...........+\sigma^{DIS}_{\nu N}
 \end{equation}

Here, $\nu$ refers neutrino, N refers nucleon, $\sigma^{tot}_{\nu N}$ refers the sum of all cross-sections, $\sigma^{QE}_{\nu N}$  refers to the cross-section for QE scattering, $\sigma^{1\pi}_{\nu N}$ refers the cross-section for single pion generation, and so on.

Neutrinos interact with matter through the exchange of $W^\pm$ and $Z^0$ bosons. At low neutrino energies, the QE scattering process is favoured. As neutrino energy rises, RES and then DIS processes become more important as seen in Figure 2 for both neutrino and anti-neutrino. Around 2.5 GeV, the DUNE flux peaks and the RES and DIS cross-sections are nearly similar at this energy. RES events can have signals that are indistinguishable from DIS events in a detector, which is a problem in practice. As a result, it's difficult to measure each stage separately.

\begin{figure*}
	\vskip1mm
	\includegraphics[width=\column]{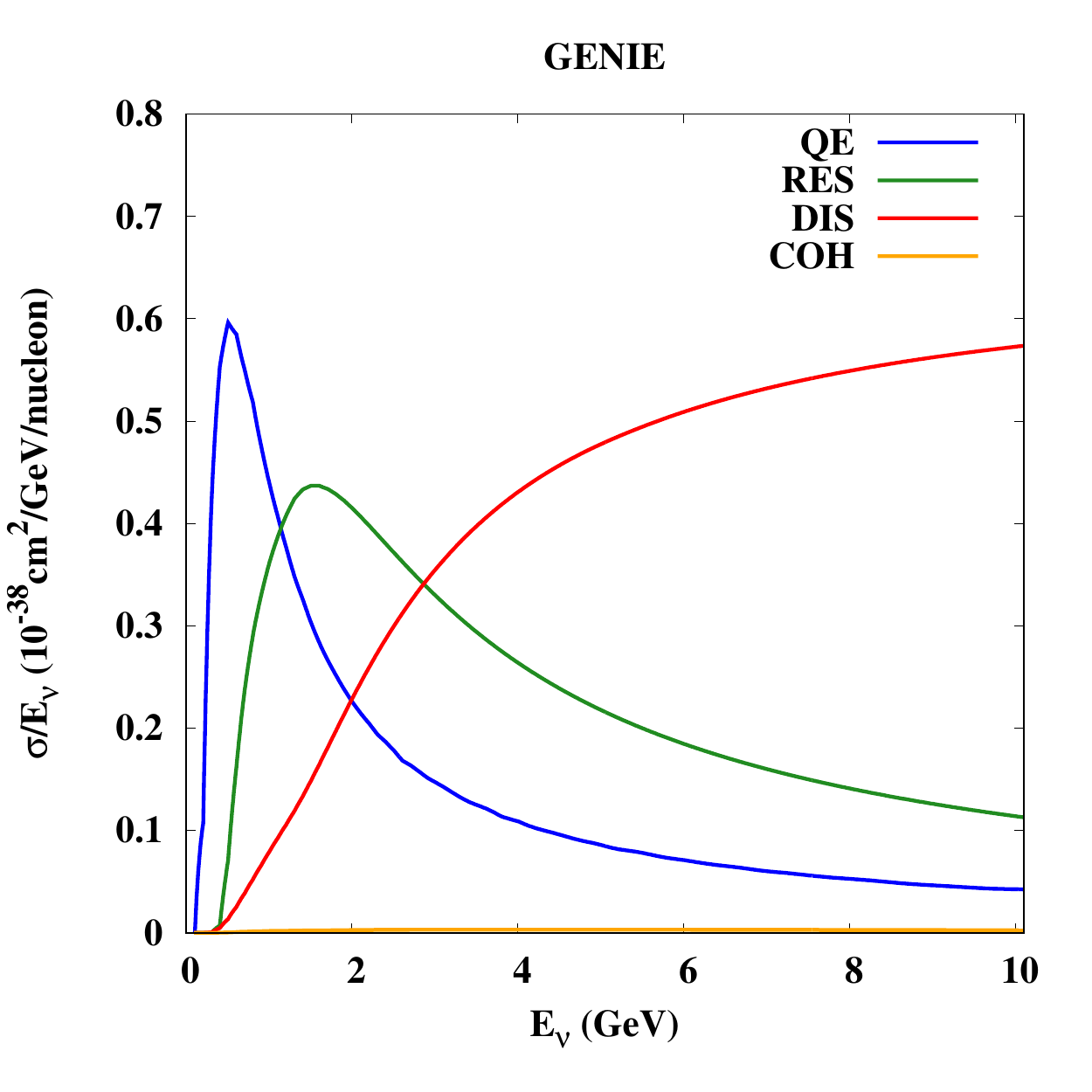}
	\includegraphics[width=\column]{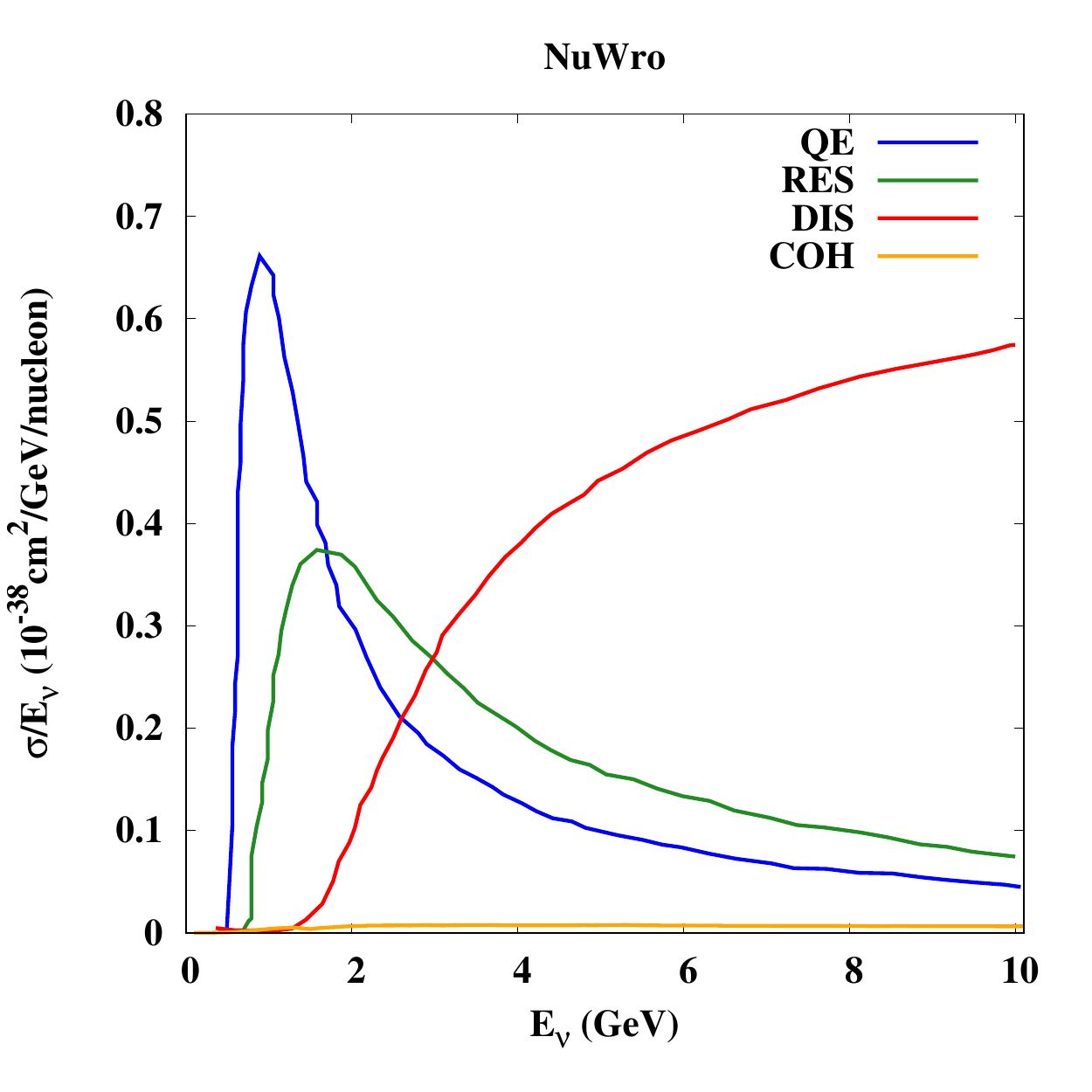}
	\includegraphics[width=\column]{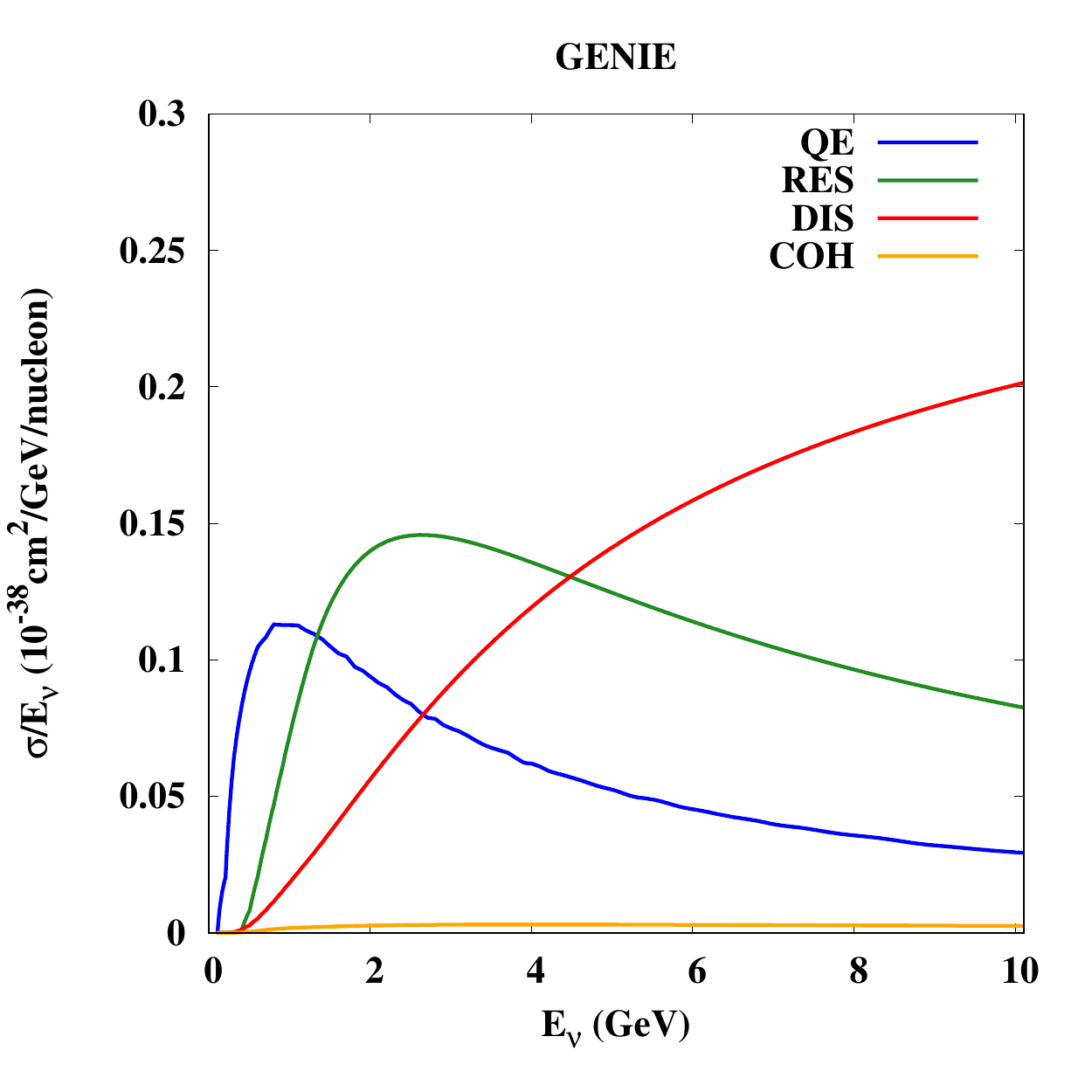}
	\includegraphics[width=\column]{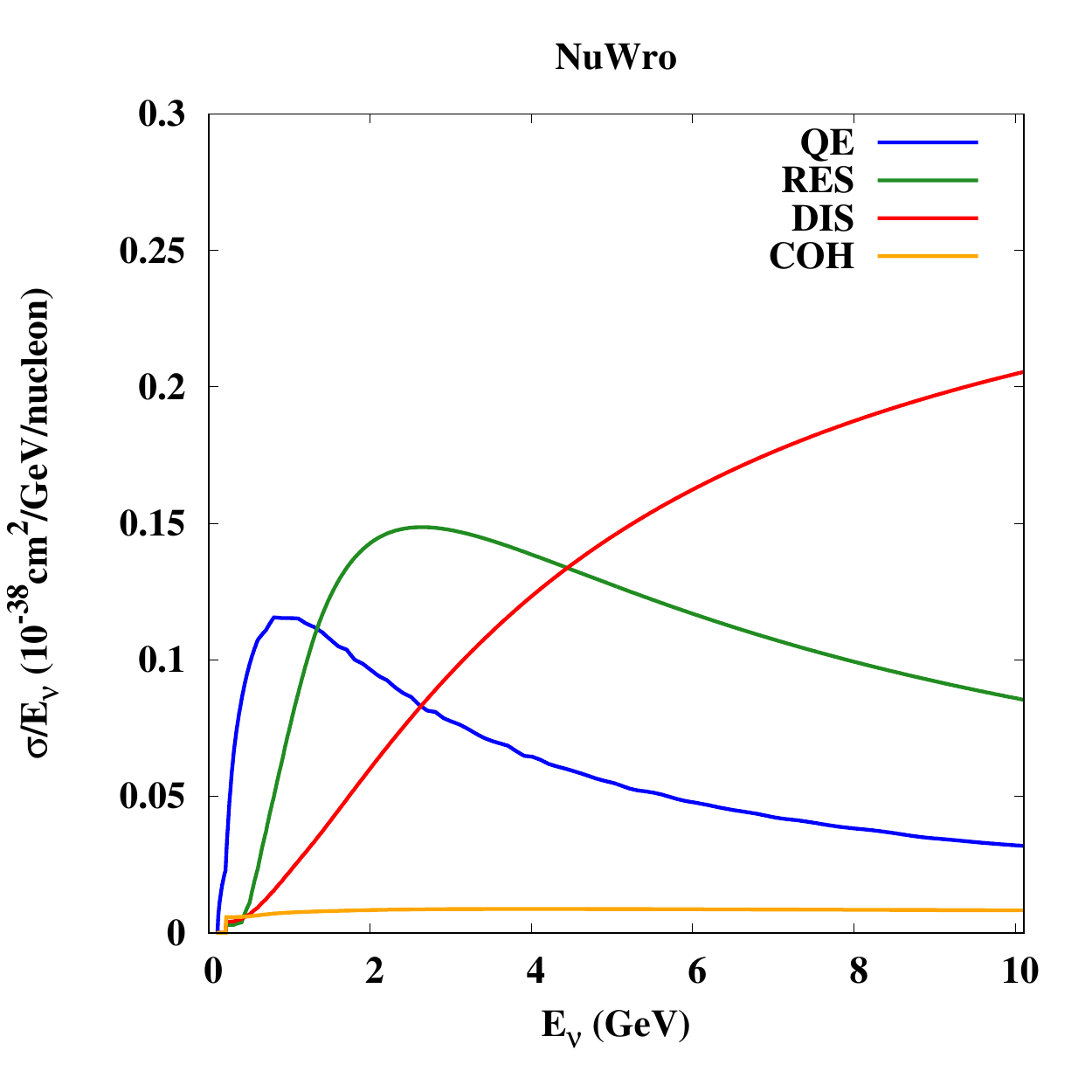}
	\vskip-3mm\caption{$\nu_\mu$-Ar (top panel) and $\bar{\nu}_\mu$-Ar (bottom panel)  interaction cross-section per nucleon as a function of neutrino energy by GENIE	(left panel) and NuWro (right panel) in the energy regime 1-10 GeV, for different charged current processes considered in our work.}

	\label{fig7}  
\end{figure*}

In QE scattering, the target nucleon remains a single nucleon in the final state, only changing its charge in CC weak interactions. This scattering does not make pions directly, but it can produce them through final state interactions. Inside the nuclear environment, hadrons can be spread elastically or inelastically, absorbed or charged exchanged, and even produce more pions. As a result, only a small percentage of events with no pions in the primary state is expected to produce pions in the final state.
For the $\nu_\mu$ beam, the CC QE scattering reaction is written as:

\begin{equation}
\nu_{\mu}+n \longrightarrow \mu^- +p
\end{equation}

\begin{figure}[htb!]
\vskip1mm
\includegraphics[width=\column]{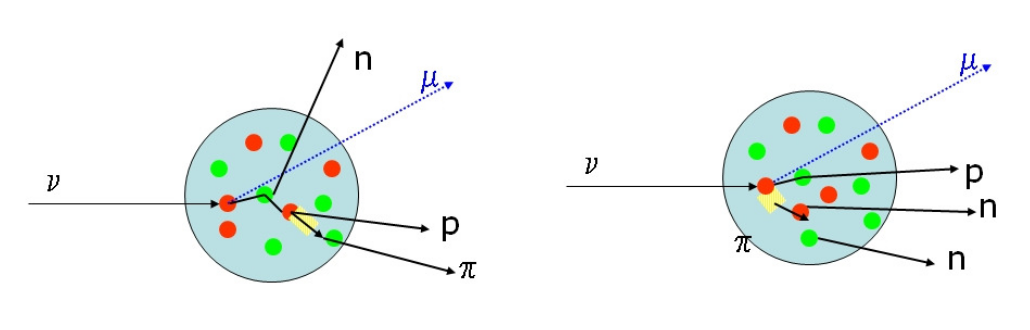}
\vskip-3mm\caption{Pion production in QE scattering (left) and absorption in RES scattering
	(right) \cite{genie}.}
\end{figure}

\begin{figure}[htb!]
	\vskip1mm
	\includegraphics[width=\column]{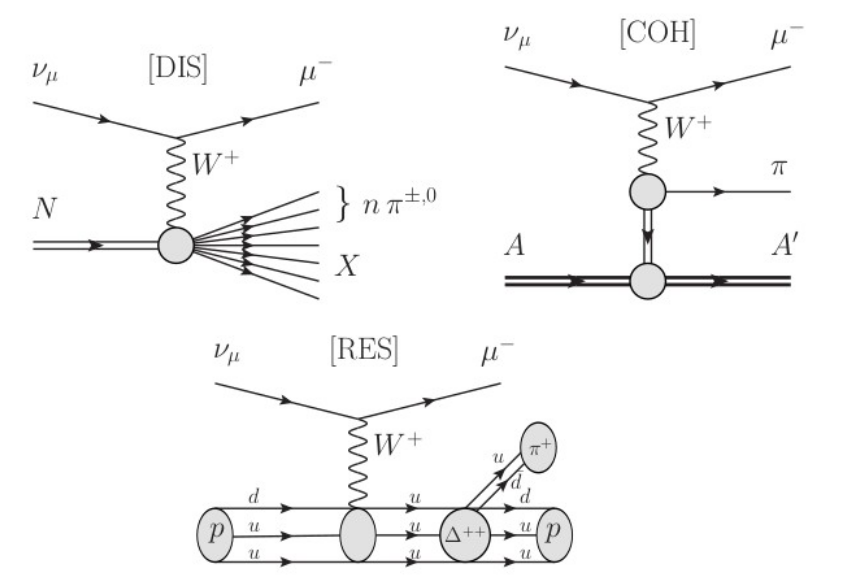}
	\vskip-3mm\caption{Pion production via various processes \cite{antonello}.}
\end{figure}

Figure 3 illustrates how pions are produced in the QE scattering process (left) and how a pion could be absorbed (right) inside the nucleus during its intranuclear journey. The most prevalent processes that directly produce pions are DIS, RES, and COH. These processes are depicted in Figure 4.

In RES scattering, resonances produce ions. In the RES production process, a neutrino excites the target nucleon to a resonance state. The resonance state that is formed quickly decays into a single nucleon and pion state. The $\Delta$(1232) resonance state contributes the most to this process, but higher resonance levels can also be generated. The CCRES scattering processes (for $\nu_{\mu}$ beam) are as follows:

\begin{equation}
\nu_\mu + p \longrightarrow \mu^- + \Delta^{++} \hspace{1.5mm};\hspace{1.5mm} \Delta^{++} \longrightarrow p+ \pi^+
\end{equation}

\begin{equation}
\nu_\mu +n \longrightarrow \mu^- + \Delta^+ \hspace{1.5mm}; \hspace{1.5mm} \Delta^+ \longrightarrow n+ \pi^+
\end{equation}

\begin{equation}
\nu_\mu +n \longrightarrow  \mu^- +\Delta^+ \hspace{1.5mm}; \hspace{1.5mm} \Delta^+ \longrightarrow p+ \pi^0 
\end{equation}

When a pion created in a RES process is absorbed in the nucleus, establishing whether the process is RES becomes difficult. As a result, the staged event is referred to as a fake event because it gives the appearance of a separate procedure. The particles recorded by the detector are like those formed at the primary vertex in a specific interaction channel.

A high-energy neutrino penetrates deep into the nucleon and scatters off a quark via the exchange of $W^{\pm}$ and $Z^0$ bosons, resulting in a lepton and a hadronic system in the ultimate state of DIS. The CC $\nu_{\mu}$ interaction process is as follows:

\begin{equation}
\nu_\mu + N \longrightarrow \mu^- +n\pi^{\pm} + X
\end{equation}

Where N is a nucleon (proton or neutron), n denotes a number, and X denotes any group of final nucleons.

Neutrinos can interact coherently with the entire nucleus, resulting in pion generation (coherent (COH) pion production). The approach for CC $\nu_{\mu}$ interaction is as follows:

\begin{equation}
\nu_\mu + A \longrightarrow \mu^- + A'+m^+
\end{equation}
Where A is the nucleus in its initial state, $A'$ is the nucleus in its final state, and $m^+$=$\pi^+$, $k^+$, $\rho^+$....

\section{Event Generators}

Neutrino event generators are simulation tools used in neutrino physics research, and they can be improved utilising experimental data from past studies. Generators serve as a link between theoretical and experimental frameworks. GENIE (Generates Events for Neutrino Interaction Experiments) and NuWro are the two neutrino event generators employed in this study (developed by Wroclaw University). In our simulations, we employed GENIE version 3.00.06 and NuWro version 19.01, which are the most recent stable releases. 

	\subsection{GENIE}
	
Cross-section models, hadronization models, and nuclear physics models are the three types of physics models employed in GENIE. Charged-current quasi-elastic scattering is described for cross-sections using the Llewellyn-Smith model \cite{llewellyn} and the newest BBBA07 form factors \cite{bodek}. The Rein-Sehgal model \cite{rein} was used to generate baryon resonances. The Bodek and Yang model \cite{Bodek} is utilised for DIS interactions, with low $Q^2$ modifications. For coherent pion production interactions, the Rein-Sehgal model with an updated PCAC formula \cite{Rein} is utilised. The default AGKY model is used in the hadronization procedure \cite{tena}. It provides a phenomenological description of the low invariant mass region using Koba-Nielson-Olsen (KNO) scaling \cite{koba}, before progressively switching to the PYTHIA/JETSET model at higher masses, with the transition being gradual and continuous. To account for the effect of the nuclear environment, the Fermi Gas model is employed, with modifications by Bodek and Ritchie to add nucleon-nucleon interactions. Factors such as Pauli blocking and discrepancies in free nucleon and nuclear structural function functions are also taken into account. Intranuclear hadron transport is handled by the INTRANUCLEAR/hA model.

\begin{table*}[ht]
	\noindent\caption{Occupancy of primary and final state hadronic systems for 1 Million events in GENIE(v3.00.06) for $\nu_\mu$-Ar CC interactions. Different topological groups for primary and final state systems were made on the basis of number of pions produced event-wise.}\vskip3mm\tabcolsep4.5pt
	
	\noindent{\footnotesize
		\begin{tabular}{|c || c|c|c|c|c|c|c|c|c|c|c||c|  }	\hline
			
			&   \multicolumn{11}{c||}{Primary hadronic system} &     \\  
			\cline{1-13} 
			Final	State & $0 \pi$  & $\pi^{0}$ & $\pi^{+}$ & $\pi^{-}$ & $2\pi^{0}$  & $2\pi^{+}$  & $2\pi^{-}$ & $\pi^{0}\pi^{+}$ & $\pi^{0}\pi^{-} $ & $\pi^{+}\pi^{-}$ & $\geq 3\pi$ & Total     \\ \hline \hline
			$0 \pi$      & 	\textbf{243896} &	24393 &	76207 &	237 &	180 &	485 &	0 &	1423 &	25 &	354 &	375	 & 347577

 \\ \hline
			$\pi^{0}$    & 2652 &	\textbf{77505} &	45010 &	188	& 2049 &	516 &	0	& 6959 &	49 &	337 &	1353 &	136618
   \\ \hline
			$\pi^{+}$    & 4098 &	7742 &	\textbf{222782} &	215 &	75	 & 3275 &	0 &	6424 &	6 &	1915 &	1488 &	248020
   \\ \hline
			$\pi^{-}$    & 739 &	7375 &	2881 &	\textbf{639}	 & 85 &	9 &	0 &	412 &	31 &	1903 &	610 &	14684
    \\ \hline	
			$2\pi^{0}$   &  3 &	3102 &	6912 &	71 &	\textbf{5545} &	215 &	0 &	3295 &	27 &	181 &	2251 &	21602
  \\ \hline	
			$2\pi^{+}$   & 10 &	419 &	3330 &	7 &	11 &	\textbf{7829} &	0 &	1443 &	1 &	50 &	1917 &	15017
 \\ \hline	
			$2\pi^{-}$   & 0 &	93 &	17 &	5 &	16 &	0 &	\textbf{0} &	4 &	5 &	11 &	88 &	239
 
     \\ \hline
	    $\pi^{0}\pi^{+}$ & 42 &	5600 &	11622 &	155 &	531 &	1739 &	0 &	\textbf{32756} &	2 &	1026 &	4331 &	57802

   \\ \hline	
    	$\pi^{0}\pi^{-}$ &  13 &	2969 &	1896 &	135 &	684 &	18 &	0 &	456 &	\textbf{121} &	1055 &	1525 &	8872

    \\ \hline	
		$\pi^{+}\pi^{-}$ & 32 &	10013 &	31548 &	681 &	241 &	404 &	0 &	3290 &	15 &	\textbf{12070} &	3585 &	61879

  \\ \hline	
		$\geq 3\pi$      &  95 &	3350 &	6095 &	497 &	1836 &	2295 &	0 &	9956 &	240 &	4066 &	\textbf{59260} &	87690

    \\ \hline \hline	
			Total        & 251582 &	142561 & 408300 &	2830 &	11253 &	16785 &	0 &	66418 &	520 &	22968 &	76783 &	1000000

     \\ \hline	
		\end{tabular}
	}
\label{table1}

	\end{table*}

\begin{table*}[ht]
	\noindent\caption{Occupancy of primary and final state hadronic systems for 1 Million events in GENIE(v3.00.06) for $\bar{\nu}_\mu$-Ar CC interactions. Different topological groups for primary and final state systems were made based on the number of pions produced event-wise.}\vskip3mm\tabcolsep4.5pt
	
	\noindent{\footnotesize
		\begin{tabular}{|c || c|c|c|c|c|c|c|c|c|c|c||c|  }	\hline
			
			&   \multicolumn{11}{c||}{Primary hadronic system} &     \\  
			\cline{1-13} 
			Final	State & $0 \pi$  & $\pi^{0}$ & $\pi^{+}$ & $\pi^{-}$ & $2\pi^{0}$  & $2\pi^{+}$  & $2\pi^{-}$ & $\pi^{0}\pi^{+}$ & $\pi^{0}\pi^{-} $ & $\pi^{+}\pi^{-}$ & $\geq 3\pi$ & Total     \\ \hline \hline
			$0 \pi$      & 	\textbf{280968} &	23903 &	212 &	94301 &	161 &	0 &	309 &	4 &	1058 &	359 &	269	 & 401544
			
			\\ \hline
			$\pi^{0}$    & 1376 &	\textbf{62882} &	190 &	55269	& 1318 &	0 &	282	& 21 &	5304 &	370 &	948 &	127960
			\\ \hline
			$\pi^{+}$    & 330 &	5932 &	\textbf{441} &	904 &	73	 & 0 &	6 &	14 &	225 &	1390 &	419 &	9734
			\\ \hline
			$\pi^{-}$    & 2707 &	9650 &	244 &	\textbf{240693}	 & 121 &	0 &	2603 &	1 &	4824 &	1371 &	1041 &	263255
			\\ \hline	
			$2\pi^{0}$   &  4 &	2810 &	69 &	5339 &	\textbf{3147} &	0 &	132 &	9 &	2442 & 176 &	1551 &	15679
			\\ \hline	
			$2\pi^{+}$   & 0 &	44 &	1 &	7 &	10 &	\textbf{0} &	0 &	1 &	0 &	4 &	48 &	115
			\\ \hline	
			$2\pi^{-}$   & 18 &	664 &	5 &	3661 &	15 &	0 &	\textbf{7407} &	0 &	1412 &	58 &	1786 &	15026
			
			\\ \hline
			$\pi^{0}\pi^{+}$ & 4 &	1831 &	138 &	512 &	314 &	0 &	9 &	\textbf{45} &	265 &	704 &	925 &	4747
			
			\\ \hline	
			$\pi^{0}\pi^{-}$ & 34 &	6261 &	199 &	8822 &	437 & 0 &	1484 &	7 &	\textbf{24415} &	709 &	3676 &	46044
			
			\\ \hline	
			$\pi^{+}\pi^{-}$ & 24 &	7644 &	744 &	21456 &	167 &	0 &	309 &	3 &	2036 &	\textbf{7146} &	2855 &	42384
			
			\\ \hline	
			$\geq 3\pi$      &  144 & 2821 & 364 &	3559 &	1093 &	0 &	2684  &	157 &	9130 &	2816 &	\textbf{50744} &	73512
			
			\\ \hline \hline	
			Total        & 285609 &	124442 & 2607 &	434523 &	6856 &	0 &	15225 &	262 &	51111 &	15103 &	64262 &	1000000
			
			\\ \hline

		\end{tabular}
	}
	\label{table2}
	
\end{table*} 

\subsection{NuWro}	
QE events are simulated in NuWro using the Llewellyn-Smith model with the most recent BBBA05 form-factors \cite{bradford}. Only the $\Delta(1232)$ contribution is based on the Rein-Sehgal model in RES, whereas the remainder of the resonances are based on the Adler-Rarita-Schwinger model \cite{grazyk}. Simulating coherent pion production interactions is done using the Rein-Sehgal model. This Rein-Sehgal model \cite{Rein} is not the same as the one used in RES. A cut on invariant mass of W<1.4 GeV defines the RES zone. NuWro describes DIS events using the Quark-Parton model \cite{sjostrand}. When W>1.6 GeV, the DIS contribution is enabled. The RES contribution is linearly turned off as the DIS contribution is turned on in the area 1.4 GeV<W<1.6 GeV. NuWro employs a combination of its hadronization model and the Bodek-Yang model. The Relativistic Fermi Gas (RFG) model is used to account for the impacts of the nuclear environment.

\section{Results}

	\begin{table*}[ht]
	\noindent\caption{Occupancy of primary and final state hadronic systems for 1 Million events in NuWro(v19.01.2) for $\nu_\mu$-Ar CC interactions. Different topological groups for primary and final state systems were made on the basis of number of pions produced event-wise.}\vskip3mm\tabcolsep4.5pt
	
	\noindent{\footnotesize

		\begin{tabular}{|c || c|c|c|c|c|c|c|c|c|c|c||c|  }	\hline
			
			&   \multicolumn{11}{c||}{Primary hadronic system} &     \\  
			\cline{1-13} 
			Final	State & $0 \pi$  & $\pi^{0}$ & $\pi^{+}$ & $\pi^{-}$ & $2\pi^{0}$  & $2\pi^{+}$  & $2\pi^{-}$ & $\pi^{0}\pi^{+}$ & $\pi^{0}\pi^{-} $ & $\pi^{+}\pi^{-}$ & $\geq 3\pi$ & Total     \\ \hline \hline 		
			$0 \pi$      & \textbf{242200} &	16748 &	55659 &	622 &	141 &	133 &	0 &	1533 &	10 &	759 &	66 &	317871 \\ \hline 
			
			$\pi^{0}$    & 3596 &	\textbf{84684} &	14875 &	136 &	1720 &	94 &	0 &	10810 &	122 &	486 &	774 &	117297   \\ \hline
			$\pi^{+}$    & 5790	& 3748 &	\textbf{210329} &	21 &	51 &	1722 &	0 &	9958 &	3 &	5169 &	516 &	237307   \\ \hline
			$\pi^{-}$    &  2454 &	4361 &	3172 &	\textbf{6581} &	69 &	14 &	1 &	479 &	92 &	5475 &	397	 &23095   \\ \hline
			$2\pi^{0}$   & 483 &	1823 &	966 &	7 &	\textbf{6785} &	17 &	0 &	3365 &	24 &	113 &	2520 &	16103    \\ \hline
			$2\pi^{+}$   & 94 &	120	 & 2354 &	0 &	4 &	\textbf{6095} &	0 &	2073 &	1 &	333 &	1206 &	12280   \\ \hline
			$2\pi^{-}$   &  58 &	122 &	108 &	31 &	11 &	1 &	\textbf{2} &	35 &	20 &	351 &	124 &	863 \\ \hline
			$\pi^{0}\pi^{+}$ & 350 &	1974 &	4311 &	3 &	396 &	502 &	0 &	\textbf{80723} &	5 &	1810 &	4720 &	94794 \\ \hline	
			$\pi^{0}\pi^{-}$ &  249 &	1360 &	780 &	50 &	426 &	8 &	0 &	967 &	\textbf{1126} &	1769 &	2956 &	9691 \\ \hline	
			$\pi^{+}\pi^{-}$ & 389 &	1269 &	4920 &	86 &	31 &	113 &	0 &	2942 &	23 &	\textbf{45126} &	4659 &	59558  \\ \hline

			$\geq 3\pi$      &  223	& 1577 &	2865 &	24 &	687 &	495 &	0 &	8600 &	69 &	4546 &	\textbf{92055} &	111141 \\ \hline \hline	
			Total        & 255886 &	117786 &	300339 &	7561 &	10321 &	9194 &	3 &	121485 &	1495 &	65937 &	109993	& 1000000
			\\ \hline	
		\end{tabular}
	}	
	\label{table3}
\end{table*}

\begin{table*}[ht]
	\noindent\caption{Occupancy of primary and final state hadronic systems for 1 Million events in NuWro(v19.01.2) for $\bar{\nu}_\mu$-Ar CC interactions. Different topological groups for primary and final state systems were made on the basis of number of pions produced event-wise.}\vskip3mm\tabcolsep4.5pt
	
	\noindent{\footnotesize

		\begin{tabular}{|c || c|c|c|c|c|c|c|c|c|c|c||c|  }	\hline
			
			&   \multicolumn{11}{c||}{Primary hadronic system} &     \\  
			\cline{1-13} 
			Final	State & $0 \pi$  & $\pi^{0}$ & $\pi^{+}$ & $\pi^{-}$ & $2\pi^{0}$  & $2\pi^{+}$  & $2\pi^{-}$ & $\pi^{0}\pi^{+}$ & $\pi^{0}\pi^{-} $ & $\pi^{+}\pi^{-}$ & $\geq 3\pi$ & Total     \\ \hline \hline 		
			$0 \pi$      & \textbf{296715} &	16914 &	71 &	75785 &	186 &	0 &	224 &	1 &	1818 & 777 & 68 &	392559 \\ \hline 
			
			$\pi^{0}$    & 2073 &	\textbf{59145} & 22 &	17674 &	1807 &	0 &	100 &	11 &	10292 &	404 &	512 &	92040   \\ \hline
			
			$\pi^{+}$    & 1060	& 2960 &	\textbf{578} &	2324 &	81 &	0 &	19 &	13 &	423 &	3675 &	233 &	11366   \\ \hline
			
			$\pi^{-}$    &  5312 &	4370 &	6 &	\textbf{259892} &	104 &	0 &	2419 &	0 &	10815 &	4111 &	498	 & 287527   \\ \hline
			
			$2\pi^{0}$   & 259 &	1081 &	1 &	795 &	\textbf{5722} &	0 &	12 &	1 &	2712 &	75 &	1840 &	12498    \\ \hline
			
			$2\pi^{+}$   & 8 &	50	 & 2 &	38 &	5 &	\textbf{0} &	0 &	0 &	20 &	146 &	46 &	315   \\ \hline
			
			$2\pi^{-}$   &  64 &	158 &	0 &	2986 &	15 &	0 &	\textbf{8426} &	0 &	2764 &	311 &	1305 &	16029 \\ \hline
			
			$\pi^{0}\pi^{+}$ & 57 &	655 &	6 &	433 &	313 &	0 &	7 &	\textbf{85} &	553 &1026 & 1658 &	4793 \\ \hline	
			
			$\pi^{0}\pi^{-}$ &  198 &	1651 &	1 &	4381 &	506 &	0 &	619 &	2 &	\textbf{66307} &	1272 &	3665 &	78602 \\ \hline
			
			$\pi^{+}\pi^{-}$ & 187 &	849 &	11 &	3355 &	34 &	0 &	107 &	1 &	2262 &	\textbf{26842} & 3018 &	36666  \\ \hline

			$\geq 3\pi$      &  86	& 641 &	3 &	1599 &	494 &	0 &	564 &	7 &	5547 &	2131 &	\textbf{56533} &	67605 \\ \hline \hline	
			
			Total        & 306019 &	88474 &	701 &	369262 &	9267 &	0 &	12497 &	121 &	103513 &	40770 &	69376	& 1000000
			\\ \hline	
		\end{tabular}
	}	
	\label{table4}
\end{table*}

We have generated 1 million similar sets of events for both  $\nu_{\mu}$ and $\bar{\nu}_\mu$ GENIE and NuWro. For each generator, the number of pions created in both primary and final states was determined event by event. These simulated results are displayed in tables that indicate the topologies of pion formed in the primary and final states.
The topology of the pions created in primary neutrino-nucleus interactions is stored in a primary state.
A final state contains the topology of the pions produced after any secondary interactions (such as intra-atomic scattering or absorption) have occurred. A description of the generators' models and physics choices has previously been provided for a fair comparison of generators.

Table 1 shows the results of GENIE simulations for 1 million $\nu_{\mu}$ events and Table 2 for 1 million $\bar{\nu}_{\mu}$ events  using default axial mass parameters of $M_A^{QE}$=0.99 GeV/c$^2$ and $M_A^{RES}$=1.12 GeV/c$^2$. The occupancy of primary and final state topologies is shown in the tables. Figure 5 shows a comparative plot for primary and final state pions in GENIE (left panel) for $\nu_{\mu}$ (top left panel) and  $\bar{\nu}_{\mu}$ (bottom left panel).

Table 3 shows the results of NuWro simulations for 1 million $\nu_{\mu}$ events and Table 4 for 1 million $\bar{\nu}_{\mu}$ events  using default axial mass parameters of $M_A^{QE}$=1.03 GeV/c$^2$ and $M_A^{RES}$=0.94 GeV/c$^2$. The occupancy of primary and final state topologies is shown in the tables. Figure 5 shows a comparative plot for primary and final state pions in NuWro (right panel) for $\nu_{\mu}$ (top right panel) and  $\bar{\nu}_{\mu}$ (bottom right panel).

\begin{table}[!htp]
	\noindent\caption{Percentage of events without pion (0$\pi$), with exactly one pion (1$\pi$) and with more than one pion (>1$\pi$) for $\nu_\mu$-Ar CC interactions. Values in brackets refer to results after final state interactions.}\vskip3mm\tabcolsep4.5pt
	\noindent{\footnotesize
		\begin{tabular}{|c | c|c| }
			\hline%
			\multicolumn{1}{|c}{\rule{0pt}{5mm}\textbf{Pions}} %
			& \multicolumn{1}{|c|}{\textbf{GENIE}}%
			& \multicolumn{1}{|c|}{\textbf{NuWro}}\\[2mm]%
			\hline%
			\rule{0pt}{5mm} 
			0$\pi$  &  25.2\%  (34.8\%)  &  25.6\% (31.8\%)    \\ 
			$1\pi^0$  & 14.2\% (13.7\%)  &   11.8\% (11.7\%)  \\ 
			$1\pi^+$  & 40.8\% (24.8\%)  &  30\% (23.7\%)     \\ 
			$1\pi^-$  & 0.3\% (1.5\%)    & 0.8\% (2.3\%)   \\ 
			$1\pi$    & 55.3\% (39.9\%)  &   42.6\% (37.7\%)     \\ 
			$>1\pi$   & 19.5\% (25.3\%)  &  31.8\% (30.5\%)   \\
			[2mm]%
			\hline
		\end{tabular}
	}
	\label{table5}
\end{table}

\begin{table}[!htp]
	\noindent\caption{Percentage of events without pion (0$\pi$), with exactly one pion (1$\pi$) and with more than one pion (>1$\pi$) for $\bar{\nu}_\mu$-Ar CC interactions. Values	in brackets refer to results after final state interactions.}\vskip3mm\tabcolsep4.5pt
	\noindent{\footnotesize
		\begin{tabular}{|c | c|c| }
			\hline%
			\multicolumn{1}{|c}{\rule{0pt}{5mm}\textbf{Pions}} %
			& \multicolumn{1}{|c|}{\textbf{GENIE}}%
			& \multicolumn{1}{|c|}{\textbf{NuWro}}\\[2mm]%
			\hline%
			\rule{0pt}{5mm} 
			0$\pi$  &  28.6\%  (40.1\%)  &  30.6\% (39.3\%)    \\ 
			$1\pi^0$  & 12.4\% (12.8\%)  &  8.8\% (9.2\%)  \\ 
			$1\pi^+$  & 0.3\% (1.0\%)  &  0.1\% (1.1\%)     \\ 
			$1\pi^-$  & 43.4\% (26.3\%)    & 36.9\% (28.9\%)   \\ 
			$1\pi$    & 56.1\% (40.1\%)  &   45.8\% (39.1\%)     \\ 
			$>1\pi$   & 15.3\% (19.8\%)  &  23.6\% (21.5\%)   \\
			[2mm]%
			\hline
		\end{tabular}
	}
	\label{table6}
\end{table}

\begin{table}[htbp]
	
	\noindent\caption{Percentage of events with single pion or no pion in the final state if there was a single pion in the primary state for  $\nu_\mu$-Ar CC interactions.}\vskip3mm\tabcolsep4.5pt
	\noindent{\footnotesize
		
		\begin{tabular}{|c | c|c|  }
			\hline
			\multicolumn{1}{|c}{\rule{0pt}{5mm}\textbf{Process}} %
			& \multicolumn{1}{|c|}{\textbf{GENIE}}%
			& \multicolumn{1}{|c|}{\textbf{NuWro}}\\[2mm]%
			\hline%
			\rule{0pt}{5mm} 
			$\pi^0\longrightarrow \pi^0$  &  54.4\%    &  71.9\%     \\ 
			$\pi^+\longrightarrow \pi^+$  & 54.6\%   &   70\%  \\ 
			$\pi^0\longrightarrow 0\pi's$  & 17.1\%   &  14.2\%     \\ 
			$\pi^+\longrightarrow 0\pi's$  & 18.7\%    & 18.5\%    \\ 
			$\pi^0\longrightarrow \pi^+$    & 5.4\%   &   3.2\%      \\ 
			$\pi^0\longrightarrow \pi^-$   & 5.2\%   &  3.7\%   \\
			$\pi^+\longrightarrow \pi^0$   & 11\%   &  5\%    \\
			[2mm]%
			\hline
		\end{tabular}
	}
	\label{table7}
\end{table} 

\begin{table}[htbp]
	\noindent\caption{Percentage of events with single pion or no pion in the final state if there was a single pion in the primary state for $\bar{\nu}_\mu$-Ar CC interactions. }\vskip3mm\tabcolsep4.5pt
	\noindent{\footnotesize

		\begin{tabular}{|c | c|c|  }
			\hline
			\multicolumn{1}{|c}{\rule{0pt}{5mm}\textbf{Process}} %
			& \multicolumn{1}{|c|}{\textbf{GENIE}}%
			& \multicolumn{1}{|c|}{\textbf{NuWro}}\\[2mm]%
			\hline%
			\rule{0pt}{5mm} 
			$\pi^0\longrightarrow \pi^0$  &  50.5\%    &  66.9\%     \\ 
			$\pi^-\longrightarrow \pi^-$  & 55.4\%   &   70.4\%  \\ 
			$\pi^0\longrightarrow 0\pi's$  & 19.2\%   &  19.1\%     \\ 
			$\pi^-\longrightarrow 0\pi's$  & 21.7\%    & 20.5\%    \\ 
			$\pi^0\longrightarrow \pi^+$    & 4.8\%   &   3.3\%      \\ 
			$\pi^0\longrightarrow \pi^-$   & 5.2\%   &  4.9\%   \\
			$\pi^-\longrightarrow \pi^0$   & 12.7\%   &  20.5\%    \\
			[2mm]%
			\hline
		\end{tabular}
	}
	
	\label{table8}
\end{table}

\begin{figure*}[hbt!]
	\vskip1mm
	\includegraphics[width=\column]{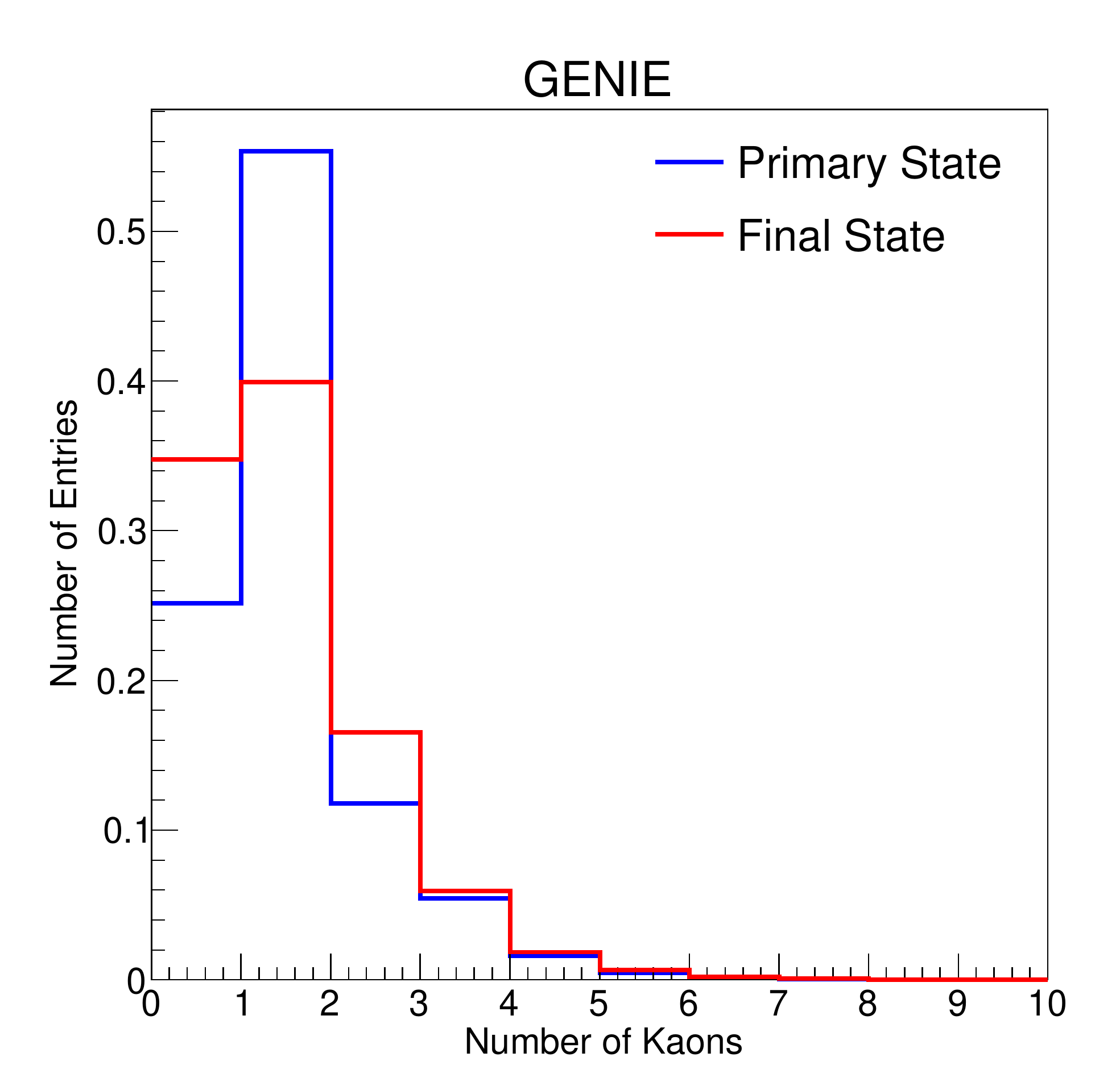}
	\includegraphics[width=\column]{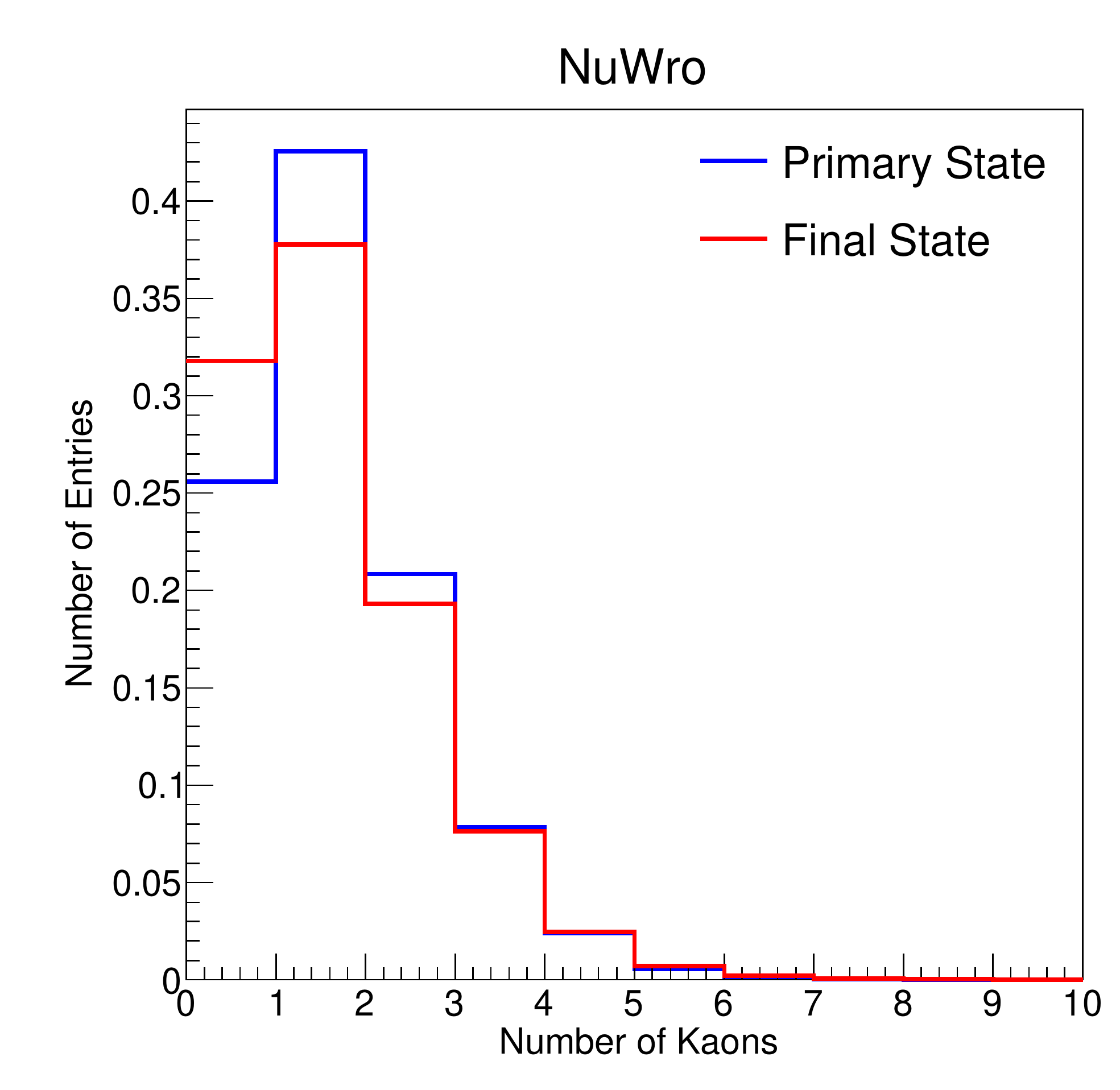}
	\includegraphics[width=\column]{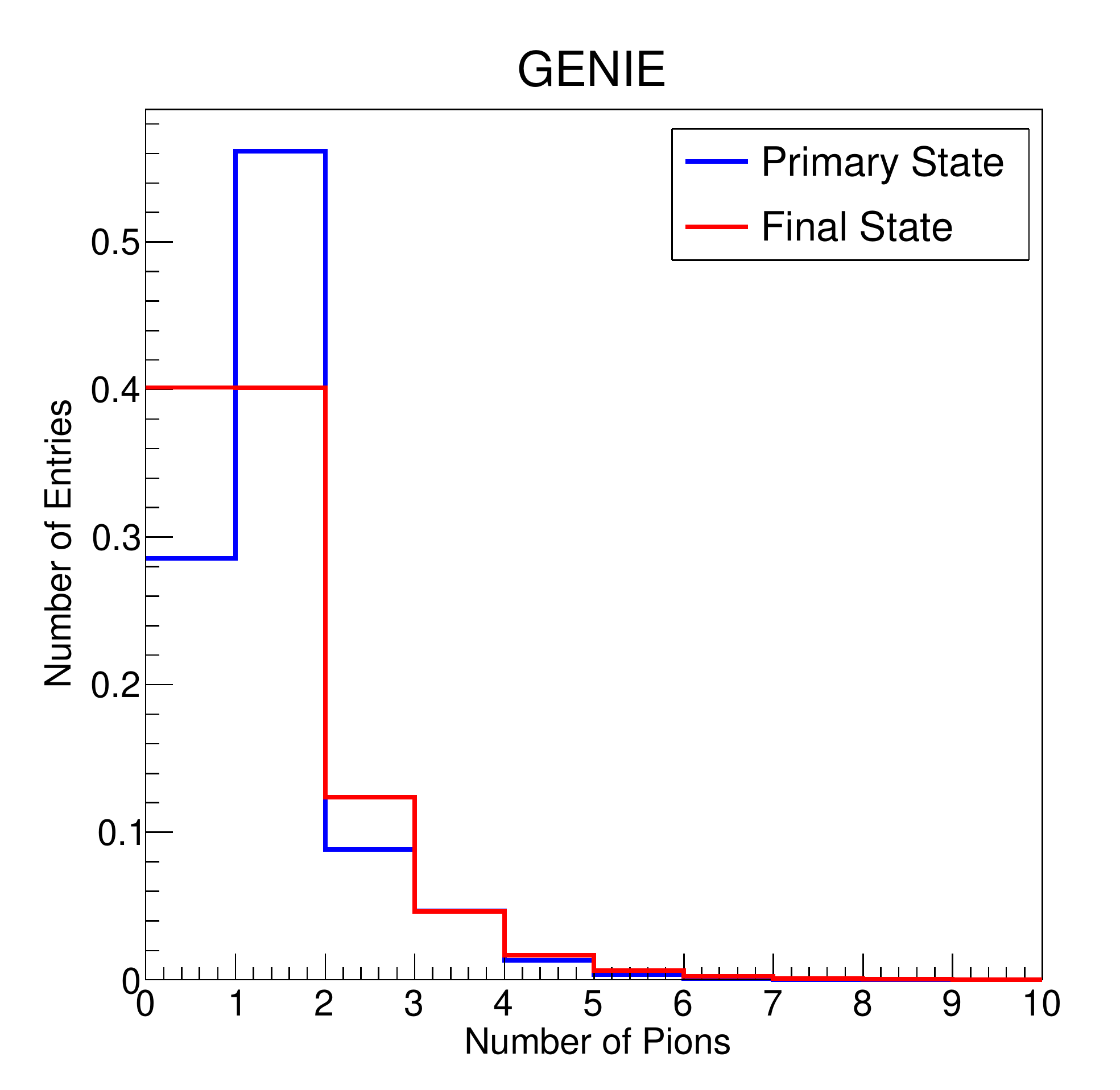}
	\includegraphics[width=\column]{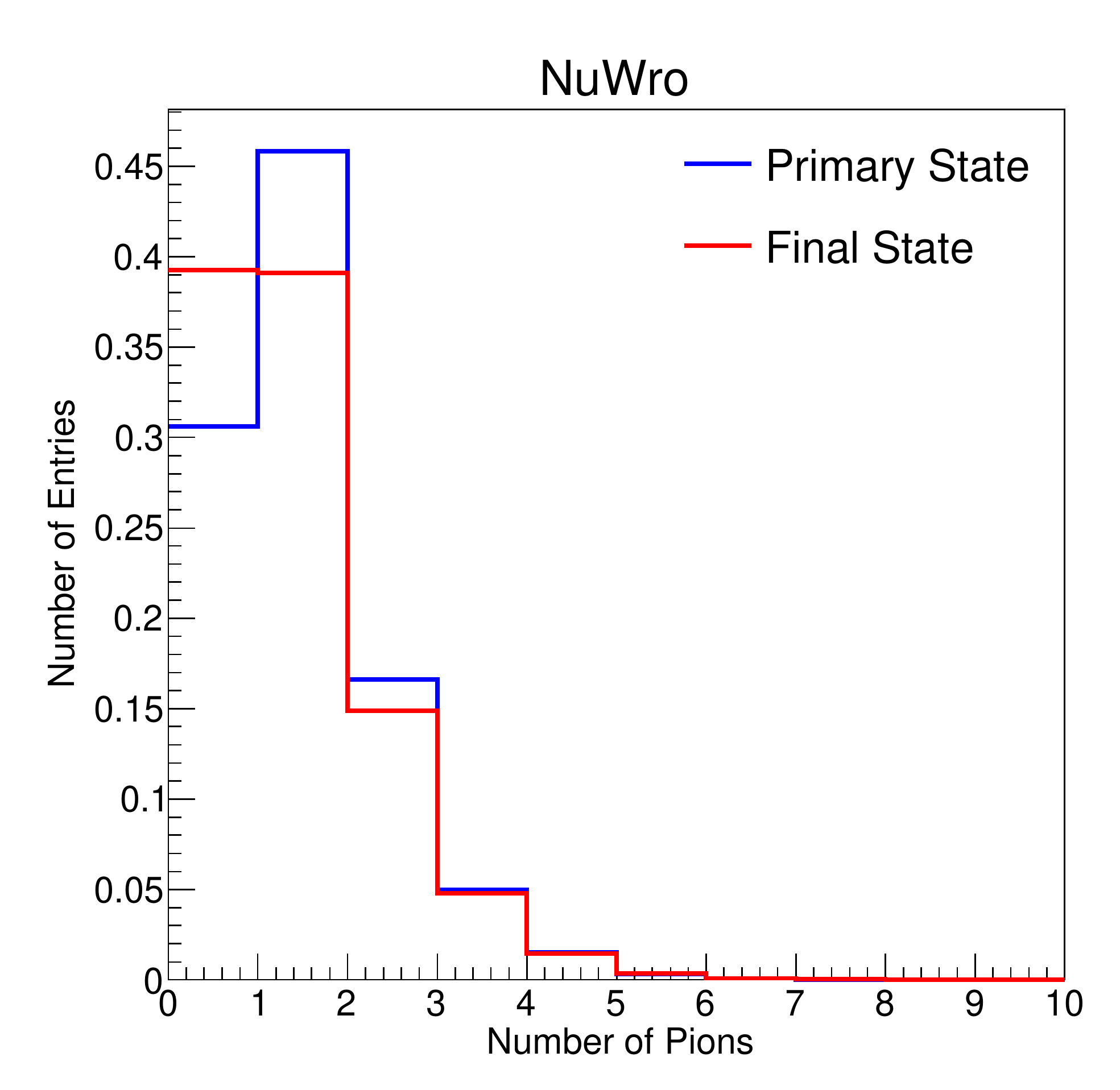}		
	\vskip-3mm\caption{Number of pions produced on event by event basis in primary and final states for GENIE (left panel) and NuWro (right panel) generators. Blue lines represent primary states and red lines represent final states in both the panels  }
	\label{fig:1}
\end{figure*}

In  case of $\nu_{\mu}$, Tables 1 (GENIE) and 3 (NuWro) illustrate that for two generators, there are considerable differences in both primary and final state topologies (i.e. the number of pions seen in primary and final states). Figure 5 (top panel) provides plots for each generator's number of pions observed on an event-by-event basis in the main and final states, revealing the disparities. Frequently, these differences are higher than statistical fluctuations. The number of  $\pi^+$  observed in the final state corresponding to  $\pi^0$  in the primary state for GENIE is 7742, but the number for NuWro is 3748, a difference of more than 50\%. The number of  $\pi^0$ observed in the final state matching  $\pi^+$  in the primary state for GENIE is 45010, but the figure for NuWro is 14875, which is a significant difference. The number of  $\pi^-$ observed in the final state equivalent  $\pi^-$  in the primary stage for GENIE is 639, whereas the figure for NuWro is 6581. The reason for these discrepancies is that DUNE flux peaks near 2.5 GeV, but QE, RES, and DIS processes all contribute significantly to the overall cross-section in this energy region. Although the models used to explain these processes are commonly shared among generators, there are numerous differences in how each generator views the merging of relative impact in this energy zone. When this effect is paired with other input parameters, visible differences across models can emerge.

In  case of $\bar{\nu}_{\mu}$, Tables 2 (GENIE) and 4 (NuWro) illustrate that for two generators, there are considerable differences in both primary and final state topologies (i.e. the number of pions seen in primary and final states). Figure 5 (bottom panel) provides plots for each generator's number of pions observed on an event-by-event basis in the main and final states, revealing the disparities. The number of  $\pi^+$  observed in the final state corresponding to  $\pi^0$  in the primary state for GENIE is 5932, but the number for NuWro is 2960, a difference of more than 50\%. The number of  $\pi^0$ observed in the final state matching  $\pi^-$  in the primary state for GENIE is 55269, but the figure for NuWro is 17674, which is a significant difference.

The final states of both generators have a higher number of zero pion ($0\pi$) topologies than the primary states, according to the tables in both the cases. This means that pions can be absorbed more easily than they can be created through intranuclear transit.
Topologies with pions in the primary and final states are more likely to be produced by RES and DIS processes, whereas topologies with $0\pi$ in the primary and final states are more likely to be produced by QE processes. Furthermore, the generators' topologies are almost identical. Changes in nuclear models and form-factor factors utilised in each generator could explain the differences.

Using Tables 1 and 3 for $\nu_{\mu}$ and  Tables 2 and 4 for $\bar{\nu}_{\mu}$  illustrates the fraction of events with no pion (0$\pi$), exactly one pion (1$\pi$), and more than one pion (>1$\pi$). Single pion production (1$\pi$), however, is recommended in GENIE, whereas multiple pion production is preferred in NuWro.

Using  Tables 1 and 3, Table 5 illustrates the fraction of events with no pion (0$\pi$), exactly one pion (1$\pi$), and more than one pion (>1$\pi$) in the case of $\nu_{\mu}$. Using  Tables 2 and 4, Table 6 illustrates the fraction of events with no pion (0$\pi$), exactly one pion (1$\pi$), and more than one pion (>1$\pi$) in the  case of  $\nu_{\mu}$. Single pion production (1$\pi$), however, is recommended in GENIE, whereas multiple pion production is preferred in NuWro.

Tables 1, 2, 3,  and 4 reveal important information about final-state interactions. To collect this data, we produced a summary table (Table 6 for $\nu_{\mu}$ and Table 8 for $\bar{\nu}_{\mu}$). The summary table depicts the topology-changing effect of intranuclear hadron transit. Out of all events having a certain primary state topology, we present the fraction of occurrences with both primary and final state topologies for each generator in this table. In the first two rows, the nucleus is translucent. These rows represent the percentage of occurrences with $1\pi$ in the primary state that will still have $1\pi$ in the final state. The pion formed in the primary vertex is more likely to not re-interact, as we can see. The effect of charge exchange is shown in the next three rows, while the fraction of pions absorbed is shown in the following rows. NuWro's current version (v19.01) is significantly more transparent than GENIE's current version (v3.00.06). This could be due to a generator's sensitivity to absorption and charge exchange processes (NuWro may have too little, whereas GENIE may have too much). Despite these differences, given the complicated nature of final-state interactions, one could argue that the agreement is still good. The results of the analysed MC generators are very similar. The finding could be particularly useful because single pion events make up the majority of the background in neutrino oscillation investigations.

\section{Summary and conclusion}
In this paper, we look at the effect of final state contacts on pion production for $\nu_{\mu}$($\bar{\nu}_\mu$) -nucleus interactions on a $^{40}$Ar target in detail. The most recent versions of the GENIE and NuWro generators were used as simulation tools. Despite having comparable sets of models, two generators can have differing pion production results in the primary and final states, as seen in Tables 1, 2, 3, and 4. This could be because the two generators implement models and other input parameters differently. Both generators show essentially equal impacts of final state interactions on pions during their intranuclear trip after being produced at the primary vertex. This is seen in Tables 5 and 6. As demonstrated in Figure 5, final state interactions cause a difference between pions observed in the detector (final state pions) and pions created at the primary vertex (pions in the primary state) In addition, Tables 1, 2, 3. and 4 reveal that both generators have more $0\pi$ topologies in the final state than in the primary state, meaning that pions are more likely to be absorbed than formed during their intranuclear transit.

Our findings suggest that the ideal method for a neutrino oscillation experiment like DUNE is to have authentic correctness of nuclear models used in neutrino event generators employed for simulation, which necessitate the use of a canonical neutrino event generator. Understanding nuclear consequences necessitates a thorough understanding of hadronic physics of neutrino-nucleus interactions.

\vskip3mm \textit{Acknowledgment.}
	One of the authors, Miss Ritu Devi offers most sincere gratitude to the Council of Scientific and Industrial Research (CSIR), Government of India, for the financial support in the form of Senior Research Fellowship, file no. 09/100(0205)/2018-EMR-I.

\end{document}